\documentclass[apj]{emulateapj}
\usepackage{apjfonts}
\usepackage{graphicx}

\shorttitle{Globular cluster bimodality and galaxy assembly}
\shortauthors{Chiara Tonini}
\journalinfo{ApJ submitted}

\begin{document}

\title{The metallicity bimodality of globular cluster systems: a test of galaxy assembly and of the evolution of the galaxy mass-metallicity relation}

\author{Chiara Tonini}
\affil{Centre for Astrophysics and Supercomputing, Swinburne University
of Technology, VIC 3122, Melbourne, Australia}

\begin{abstract}

We build a theoretical model to study the origin of the globular cluster metallicity bimodality in the hierarchical galaxy assembly scenario.
The model is based on empirical relations such as the galaxy mass-metallicity relation $[\rm O/\rm H]-\rm M_{star}$ as a function of redshift, and on the observed galaxy stellar mass function up to redshift $\rm z \sim 4$. We make use of the theoretical merger rates as a function of mass and redshift from the Millennium simulation to build galaxy merger trees. We derive a new galaxy $[\rm Fe/\rm H]-\rm M_{star}$ relation as a function of redshift, and by assuming that globular clusters share the metallicity of their original parent galaxy at the time of their formation, we populate the merger tree with globular clusters. We perform a series of Monte-Carlo simulations of the galaxy hierarchical assembly, and study the properties of the final globular cluster population as a function of galaxy mass, assembly and star formation history, and under different assumptions for the evolution of the galaxy mass-metallicity relation. 
The main results and predictions of the model are the following. 1) The hierarchical clustering scenario naturally predicts a metallicity bimodality in the galaxy globular cluster population, where the metal-rich subpopulation is composed of globular clusters formed in the galaxy main progenitor around redshift $\rm z \sim2$, and the metal-poor subpopulation is composed of clusters accreted from satellites, and formed at redshifts $\rm z \sim 3-4$. 2) The model reproduces the observed relations by Peng et al. (2006) for the metallicities of the metal-rich and metal-poor globular cluster subpopulations as a function of galaxy mass; the positions of the metal-poor and metal-rich peaks depend exclusively on the evolution of the galaxy mass-metallicity relation and the $[\rm O/\rm Fe]$, both of which can be constrained by this method. In particular, we find that
the galaxy $[\rm O/\rm Fe]$ evolves linearly with redshift from a value of $\sim 0.5$ at redshift $\rm z \sim 4$ to a value of $\sim 0.1$ at $\rm z=0$.
3) For a given galaxy mass, the relative strenght of the metal-rich and metal-poor peaks depends exclusively on the galaxy assembly and star formation history, where galaxies living in denser environments and/or early types galaxies show a larger fraction of metal-poor clusters, while galaxies with a sparse merger history and/or late type galaxies are dominated by metal-rich clusters. 4) The globular cluster metallicity bimodality disappears for galaxy masses around and below $\rm M_{star} \sim 10^9 \ \rm M_{\odot}$, and for redshifts $\rm z>2$.
 
\end{abstract}


\section{Introduction}  

Globular cluster (GC) systems in galaxies have become a useful tool to study the mechanisms of galaxy formation. Thanks to a rise in the level of details in observations, now we can gain insight into the colour, metallicity and abundance gradients of such systems for a large number of galaxies, and build statistically solid scaling relations between GC and galaxy properties. 

GCs are for the most part old objects, with ages estimated to be $>10$ Gyr (Brodie et al. 2005, Strader et al. 2005, Peng et al. 2006). Therefore, not only they have survived any violent event in the assembly of their host galaxy, but they also provide a chemical record of the galaxies where they were formed (Pota et al. 2012). 
Coupled with the fact that they are very luminous, they make for excellent probes of the fossil records of galaxies and shed light on the mechanisms of galaxy assembly and star formation history.

Of particular interest is the metallicity distribution of GCs in galaxies. Galaxies of all morphologies have a GC population with an average metallicity that correlates with the galaxy stellar mass or luminosity (as first shown by Brodie $\&$ Huchra 1991; Lotz et al. 2004, Peng et al. 2006).
In addition, most galaxy GC systems exhibit a colour bimodality (Zepf $\&$ Ashman 1993, Ostrov et al. 1993, Whitmore et al. 1995, Elson $\&$ Santiago 1996, Peng et al. 2006, Spitler et al. 2006, Strader et al. 2006, Larsen et al. 2001). This is driven by a metallicity bimodality, with bluer GCs being more metal-poor and redder GCs being more metal-rich, while both populations are old ($>10$ Gyr) (Forbes et al. 2001, 1997a,b, 2011, Peng et al. 2006, Strader et al. 2005, 2006, C\^ot\'e et al. 1998, Puzia et al. 2005, Pierce et al.2006, Brodie et al. 2005, Brodie $\&$ Strader 2006). Although there is some debate in the literature (see Yoon et al. 2006), this result has been confirmed spectroscopically (Brodie et al. 2005, 2012, Cohen et al. 2003, Strader et al. 2005, Alves-Brito et al. 2011, Usher et al. 2012). The well-defined metal-rich and metal-poor GC sequences separately follow two \textit{galaxy stellar mass - GC metallicity} relations $[\rm Fe/\rm H]_{GC} - \rm M_{star}$, of which the metal-rich one is stronger and tighter, while the metal-poor one is weaker and exhibits a larger scatter (Larsen et al. 2001, Strader et al. 2006, Peng et al. 2006, C\^ot\'e et al. 1998). There is increasing evidence that these features of GC systems are universal, from giant ellipticals to dwarfs (Strader et al. 2006), although some galaxies show an even more complex situation, with multiple metallicity peaks (see for instance Peng et al. 2006, Blom et al. 2012). 

Intriguingly, the differences between blue/metal-poor and red/metal-rich GCs also extend to their dynamical properties, as shown in recent observations (Pota et al. 2012). The two subpopulations have different spatial distributions inside the host galaxies: the metal-rich GCs are more centrally concentrared, with a radial distribution profile that follows closely the spheroidal stellar component of the galaxy, while the  metal-poor GCs show a more extended distribution, and is likely physically associated with the stellar halo (Bassino et al. 2006, Goudfrooij et al. 2007, Peng et al. 2008, Forbes et al. 2012, Pota et al. 2012; the M87 data of Strader et al. 2011 represent the best example of the close spatial coupling of metal-rich GCs with galaxy starlight and the more extended distribution of  metal-poor GCs).
Correspondingly, the kinematics of the metal-rich subpopulation follows that of the main stellar component, including rotation (Strader et al. 2011), while the metal-poor subpopulation shows larger velocity dispersion and small or null net rotation.

A scenario has been proposed where GCs are formed in gas-rich (major) merger events;
at high redshift ($\rm z>4-5$), early mergers of smaller hosts produce metal-poor GCs, while later mergers of more evolved galaxies in high density environments produce metal-rich GCs (Muratov $\&$ Gnedin 2010, Kravtsov $\&$ Gnedin 2005, Bekki et al. 2007, 2008). These models however encounter a number of problems; there is no clear prediction about any metallicity bimodality or galaxy mass-GC metallicity relations, and the resulting ages of the metal-rich GCs are too young (Muratov $\&$ Gnedin 2010), an \textit{ad hoc} mechanism is needed to shut off blue/metal-poor GC formation (Bekki et al. 2008, Beasley et al. 2002), and an analysis of the observed GC abundance and metallicity gradients is not compatible this kind of formation mechanism (Arnold et al. 2011). 

Alternatively it has been proposed that, rather than originating from two main epochs or modes of GC formation, the GC chemo-dynamical bimodality can stem from the galaxy assembly history, without invoking mergers as the GC formation mechanism. In this scenario the metal-rich GC subpopulation is formed together with the bulk of the galaxy stellar component in an early violent dissipative phase, and during a later slower phase the metal-poor GC subpopulation is accreted, via minor mergers (Forbes et al. 2011, 1997a,b, Arnold et al. 2011, Masters $\&$ Ashman 2010), or via stripping of GCs from satellites (C\^ot\'e et al. 1998, 2000). The main difference with the merger scenario is that GCs of different metallicities are formed in different galaxies, and then brought together by galaxy assembly, rather than being formed in the same galaxy at different stages of the galaxy evolution. In this work we call this the "assembly scenario''.

\subsection{This work}

In this work we want to put the "assemby scenario'' in the context of the hierarchical structure formation theory, 
and investigate whether the GC metallicity bimodality indeed originates from the hierarchical nature of galaxy assembly. 
In other words, is the GC metallicity bimodality a natural \textit{prediction} of hierarchical clustering? 

To answer this question, we build a model to produce the assembly history of galaxies and their GC population, in a series of Monte-Carlo simulations.  
We base our model galaxy properties on observed scaling relations as a function of redshift, such as the galaxy mass-metallicity relation $[\rm O/\rm H]-\rm M_{star}$ relation, and the galaxy stellar mass function. We assume that galaxies at $\rm z=0$ were formed through a combination of local (in-situ) star formation and accretion of satellite galaxies in a series of merger episodes spanning the lifetime of the galaxy; the merger rates are obtained from the Millennium simulation. We populate each galaxy in the merger tree with GCs, assuming that they share the metallicity of the main stellar component of their parent galaxy at the epoch of their formation. When a satellite is accreted, so is its GC population. 

We investigate under what conditions the final GC population shows the metallicity bimodality, and follows the observed metal-rich and metal-poor \textit{galaxy stellar mass - GC metallicity} relations, as well as the observed galaxy mass - GC number abundance relation (Peng et al. 2006, 2008, Strader et al. 2006). 

The novelty of this analysis is that it provides constraints and predictions 1) on the galaxy $[\rm Fe/\rm H] - \rm M_{star}$ relation as a function of redshift, 2) on the galaxy assembly and star formation history, and 3) on the evolution of the GC bimodality, and ultimately it presents a method to test the hierarchical galaxy formation. 

This paper is organised as follows. In Sections 2 and 3 we present the model: in Section 2 we describe the galaxy assembly and the globular cluster formation, and the Monte Carlo simulation; in Section 3 we present the derivation of the fiducial galaxy mass-metallicity relation. In Section 4 we present our results for the globular cluster metallicity distribution and its implications to constrain galaxy formation, and in Section 5 we discuss our findings. Section 6 is a summary of our conclusions.

\section{The model: galaxy assembly and globular cluster formation}   

Consider a galaxy of stellar mass $\rm M_0$ at redshift $\rm z=0$.
This object represents the final stage of a \textit{merger tree}, i.e. a system of independent progenitor galaxies which were accreted and contributed to all the mass components (dark matter, gas, stars, GCs) that now characterise the galaxy. At any given time, we identify the \textit{main progenitor} in the merger tree as the most massive galaxy that is present in the tree, while we (improperly) call \textit{satellites}  the rest of the objects. For any given galaxy at $\rm z=0$, we build a Monte Carlo simulation with $N$ realisations of the merger tree, i.e. $\rm N$ different assembly histories. We performed numerical tests on $\rm N$ in a range $\rm N=[10,10^6]$, finding convergence of our results for $\rm N \geq 100$. The plots in this work are made with $\rm N=10^3$.

Galaxies evolve depending on the mass of the host dark matter halo and on the density of the surrounding environment. In the hierarchical clustering framework smaller objects virialise earlier (see for instance Frenk $\&$ White 2012), so they contain older, metal-poor stars. Their cycle of star formation and feedback is less efficient, and supernovae winds are more effective in expelling metals from the galaxy, factors that contribute to slow down the rise of metallicity in their stellar populations. At the same time, in more massive galaxies the deeper potential wells render supernovae winds less effective in expelling metals, and the enhanced ability to retain gas allows for sustained star formation and more stellar generations. As a consequence, at all redshifts a monothonic positive mass-metallicity correlation $\rm M_{star} - [\rm Fe/\rm H]$  is in place for all galaxies in the merger tree. The derivation of this relation from the observed $[\rm O/\rm H] - \rm M_{star}$ relation will be described in detail in the next Section. 

We assume the appearance of a globular cluster population in a galaxy is an event of a relatively short duration, and in general not associated with the quiescent star formation phase, but indicative of a particularly intense evolutionary phase. This assumption is sustained by a number of observations. 
First, the observed masses of GCs can reach up to $\rm M \sim 10^6 \ \rm M_{\odot}$, requiring very intense bursts of star formation.
Secondly GCs, which in general are well described by single stellar populations (SSP, i.e. coheval ensembles of stars that share the same metallicity), are for the most part old objects, with ages $>10$ Gyr (as referenced in the Introduction).
Fittingly, the observed GC ages put the epoch of their appearance squarely at the peak of the cosmic star formation history, determined to be at redshifts $\rm z \geq 2-4$ (Hopkins $\&$ Beacom 2006, Bouwens et al. 2009). 
Third, the mean metallicity of the GC population is observed to be higher in more massive galaxies, with a $\rm M_{star} - [\rm Fe/\rm H]_{GC}$ parallel to that of the galaxy mass-metallicity relation (see for instance Larsen et al. 2001, Peng et al. 2006, C\^ot\'e et al. 1998), in support of the idea that the GC population is closely related to the main stellar component and is similarly affected by halo mass and environment, i.e. more massive galaxies form their bulk of their stars later, from more enriched gas. 

We assume that the old globular cluster population (ages $>10$ Gyr) were formed in galaxies at the peak of their star formation activity, likely $\rm L^*$ galaxies, at all redshifts $\rm z>2$. Note that, following this assumption, the older globular clusters formed at higher redshifts in smaller systems, and with a lower metallicity. This is in accord with observations, that estimate metal-poor GCs to be about $1-2$ Gyr older than metal-rich GCs (Dotter et al. 2011, Puzia et al. 2005, Woodley et al. 2010), although the precision of the age measurement for extragalactic GCs is too low to confidently discriminate ages differences at this level (Strader et al. 2005). 
We assume as likely candidates for the formation sites of GCs either the massive star-forming clumps observed in high redshift ($\rm z>2$) galaxies (Shapiro et al. 2010), or the central regions of galaxies subject to episodes of violent dissipative collapse. In both cases, the clumpiness and turbulence of the gas plays a fundamental role in boosting the star formation and producing GCs, along with the galaxy main stellar component (Shapiro et al. 2010). 

The frequency of globular clusters $\rm T_N$ is defined as the number of GCs per unit galaxy mass of $10^9 \ \rm M_{\odot}$, and at $\rm z \sim 0$ it is constrained by observations (Peng et al. 2008; see also Spitler et al. 2008, Rhode et al. 2007, Rhode 2012). 
In the galaxies in the merger tree, $\rm T_N$ depends on the interplay of different factors, like the mean gas density (which depends on the depth of the galactic potential well), the metallicity, the feedback regime, and the competing 'regular' star formation that feeds the main stellar component. In lack of other observational constraints, we assume that the redshift $\rm z \sim 0$ observed relation $\rm T_N \ - \rm M_{star}$ holds at all redshifts, so that the total number of local GCs that each galaxy produces is $\rm N_{GC} = \rm T_N (\rm M_{star}) \times \rm M_{star}$. In addition, galaxies below $\rm M_{star} = 10^9 \rm M_{\odot} = \rm M_{min}$ do not form globular clusters, consistently with the observed $\rm T_N \ - \rm M_{star}$ relation (Peng et al. 2008), which yields $\rm N_{GC} < 10$ for $\rm M_{star} \sim 10^9 \rm M_{\odot}$ (see also Muratov $\&$ Gnedin, 2010). We also assume that, once formed, the local GCs stabilise themselves in dynamical equilibrium with the galaxy, and therefore remain kinematically coupled with the main stellar component.

We follow the evolution of the merger tree from redshift $\rm z_{in}$, when the galaxy main progenitor forms its local globular clusters. We assume the epoch is 
$\rm z_{in} \sim 2$. The main progenitor is likely to have a rich gas component and is near the peak of its star formation history; it has a stellar mass $\rm M_1$ and a mean total metallicity $\rm Z_1$, which follows the galaxy mass - metallicity relation $\rm M_{star} - [\rm Fe/\rm H]$ at $\rm z \sim \rm z_{in}$. 
In each Monte-Carlo realisation, we assume that the metallicity of the locally formed GCs is peaked aroung $\rm Z_1$, with a gaussian distribution with $\sigma=0.2$ (consistent with C\^ot\'e et al. 1998, Bekki et al. 2008), which takes into account a non-instantaneous mixing of the metals, and the fact that the GC formation covers a short but finite time-span, in which the mean galaxy metallicity can vary. The number of local GCs in the main progenitor is $\rm N_{GC} = \rm T_N (\rm M_1) \times \rm M_1$.

The main progenitor is the most massive galaxy in the merger tree, and is the last one to have its globular cluster population in place. At this point in time, the satellites in the merger tree, which by definition have masses $\rm M_i<\rm M_1$, have already formed their own GCs, and we put such epoch of formation around $\rm z \sim 3-4$ (see also Shapiro et al. 2010). Each satellite metallicity follows the galaxy $\rm M_{star} - [\rm Fe/\rm H]$ at that epoch, and in each satellite of mass $\rm M_i$ the GC metallicity is peaked around the current galaxy mean metallicity $\rm Z_i$, while the number of GCs is $\rm N_{GCi} = \rm T_N (\rm M_i) \times \rm M_i$. 
We assume a gaussian distribution of the GC metallicities in each satellite, peaked around $\rm Z_i$ with $\sigma=0.3$, which takes into account the combined effect of the scatter in the GC metallicity in each satellite ($\sim 0.2$), plus an additional uncertainty ($\sim 0.2$) due to the scatter in the star formation histories of satellites (sensitive to environment for instance), which affect the satellite's metallicity and the exact epoch of GC formation.

The main progenitor $\rm M_1$ evolves into the $\rm z=0$ galaxy $\rm M_0$ through two main channels: by accreting stellar mass in the form of satellites, and by forming stars locally. If we define as $\rm M_{SF}$ the mass in stars that are formed inside the main progenitor at any time \textit{after} the GC formation (including merger-triggered star formation), then the stellar mass accreted from satellites is $\rm M_{sat}=\rm M_0-\rm M_1-\rm M_{SF}$.  $\rm M_{sat}$ is the sum of the stellar mass present in all satellites at redshift $\rm z_{in}$, under the assumption that the satellite $\rm T_N$ remains constant (i.e. satellites do not have a prolongued star formation history aftet the GC formation). 
The ratios $\rm M_1/\rm M_0$ and $\rm M_{SF}/\rm M_0$ are free parameters in the model, and they constrain the assembly and star-formation history. 

For each galaxy characterised by ($\rm M_0, \rm M_1, \rm M_{SF}/\rm M_1$), we run a Monte-Carlo simulation of $\rm N$ realisations of the galaxy merger history, from $\rm z_{in}$ to $\rm z=0$. In each run, we randomise the metallicity distribution of the main progenitor's GCs around $\rm Z_1$. 
We build the merger tree based on the observed stellar mass function (SMF) of Marchesini et al. (2009), and the theoretical merger rates obtained from the Millennium simulation (Springel et al. 2005, Fakhouri et al. 2010). 
In each realisation, we randomise both the mass of the accreted satellites and the redshift of accretion; after sampling a random redshift in the interval [$\rm z \sim 4 - 0$], we interpolate the observed stellar mass function to that redshift, and we sample a random satellite mass from it with an acceptance-rejection algorithm. 
This provides us with a series of merger candidates; each of them is weighted with the mean merger rate, which represents the probability for a merger to happen, given the mass of the main progenitor $\rm M$, the ratio between the masses of the satellite and the main progenitor $\epsilon$, and the redshift: 

\begin{equation}
\frac{\rm d \rm N_m}{\rm d\epsilon \rm d\rm z} (\rm M, \epsilon, \rm z) = \rm A \left( \frac{\rm M}{10^{12} \rm M_{\odot}} \right)^{\alpha} \epsilon^{\beta} \  \rm{exp} \left[ \left( \frac{\epsilon}{\epsilon_0} \right)^{\gamma} \right] (1+\rm z)^{\eta} ~, 
\end{equation}

where the best-fit parameters are characterised as $(\alpha,\beta,\gamma,\eta) = (0.133, -1.995, 0.263, 0.0993)$ and $(\rm A,\epsilon_0)=(0.0104, 9.72 \times 10^{-3} )$ (Fakhouri et al. 2010). From Fig. (1) in Fakhouri et al. (2010) it is evident that the merger rate increases with increasing redshift and decreasing halo mass, and hugely favours small ratios $\epsilon << 1$ between the satellite's and the main progenitor's masses. 

At each timestep in our merger history, we add the weighted mass of each satellite to the stellar mass of the main progenitor $\rm M$, which grows in time, and we continue until the total accreted mass is equal to $\rm M_{sat}$. Each merged satellite carries a population of $\rm N_{GCi} = \rm T_N (\rm M_i) \times \rm M_i$  globular clusters,  with metallicity centered around $\rm Z_i$ and randomised in each run.
This completes one realisation in the Monte Carlo simulation and represent one of the $\rm N$ merger histories that we build for each galaxy. 
For each merger history, we obtain a total GC metallicity distribution that is given by the superposition of the contributions from the main progenitor and all the satellites. After $\rm N$ realisations, we produce a mean of the total GC metallicity distribution. 

In addition, we also explore a scenario where new globular clusters can be created in gas-rich merger events. In this case, we consider that at a redshift $\rm z_{new}$ a merger event is characterised by a gas mass $\rm M_{gas}$ that is turned into stars and globular clusters, producing $\rm N_{new}$ new globular clusters of random metallicity peaked around $\rm Z_{new}$ (the metallicity of the gas), with a gaussian distribution with $\sigma=0.2$.

For a given final galaxy mass, the model uses 2 free parameters. The ratio $\rm M_1/\rm M_0$ between the mass of the main progenitor at the epoch when it forms its local GCs and the final stellar mass of the galaxy is the \textit{assembly parameter}; the ratio $\rm M_{SF}/\rm M_0$ between the mass of the stars formed locally in the evolving main progenitor \textit{after} the GC formation and the final stellar mass of the galaxy is the \textit{star formation history parameter}.

\section{The model: evolution of the galaxy mass-metallicity relation and globular cluster metallicity}  

The main source of systematic uncertainty in the model comes from the redshift evolution of the galaxy mass-metallicity relation $\rm M_{star} - [\rm Fe/\rm H]$.
Although this is in principle constrained by observations, we feel that there currently is a lack of consensus on the evolution of $\rm M_{star} - [\rm Fe/\rm H]$ at the level of precision required for this investigation. 
For this reason, to assign a metallicity to the galaxies in the merger tree and their globular cluster systems, we build a fiducial $\rm M_{star} - [\rm Fe/\rm H]$ relation as a function of redshift, and we explore the consequences of varying this relation on the model. As a sanity check, to obtain the metallicity of the satellite GCs we also make use of the \textit{total} $\rm M_{star} - [\rm Fe/\rm H]_{GC}$ relation of Peng et al. (2006). 

\begin{figure} 
\includegraphics[scale=0.45]{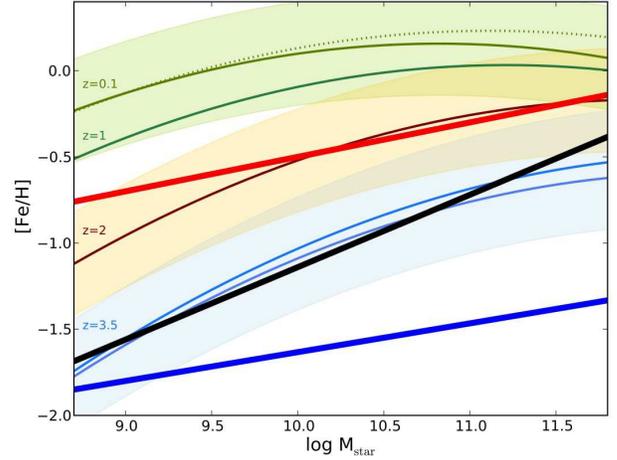}  
\caption{The evolution of the galaxy mass-metallicity relation and the globular cluster metallicity relations from Peng et al. (2006). \textit{Thick lines: black} average total GC relation, \textit{red}: metal-rich GC relation, \textit{blue}: metal-poor GC relation. 
\textit{Thin lines with shaded areas}: the evolution of the galaxy mass-metallicity relation as obtained in this work (see text), with the $1\sigma$ regions ($\rm z=1$ region omitted for clarity). The $\rm z=0.1$ \textit{dotted line} is the relation we obtained from Tremonti et al. (2004). All the \textit{solid lines} represent the relation we obtain from Maiolino et al. (2008). For the $\rm z=3.5$ relation we show the results for both Maraston (2005) and Bruzual $\&$ Charlot (2003) stellar population models.  }
\label{mets}
\end{figure}

The GC metallicity as a function of galaxy mass is provided in terms of $[\rm Fe/\rm H]$, which is a proxy for the total metallicity (Fig. 14 of Peng et al. 2006, Shapiro et al. 2010). The galaxy metallicity on the other hand is often measured in terms of the quantity $12+\rm log(\rm O/\rm H)$; in particular we consider the relations provided by Maiolino et al. (2008) up to $\rm z \sim 3.5$ for the AMAZE (Assessing the Mass-Abundance redshift[-Z] Evolution) program. In the lower redshift bin, this relation is consistent with the one provided by Tremonti et al. (2004) for a sample of $~53000$ galaxies in the Sloan Digital Sky Survey. The Maiolino relations can be parameterised as follows: 

\begin{equation}
12+\rm log(\rm O/\rm H)=-0.0864 (\rm log \rm M_{star}-\rm log \rm M_0)^2+\rm k_0~,
\end{equation}

and for redshifts $\rm z=(0.07,0.7,2.2,3.5)$ the parameters are $\rm M_0=(11.18,11.57,12.38,12.76/12.87)$ and $\rm k_0=(9.04,9.04,8.99,8.79/8.9)$ (Maiolino et al. 2008). 
To obtain a $[\rm Fe/\rm H]$ estimate from the quantity $\rm log(\rm O/\rm H)$ we need to establish the $12+\rm log(\rm O/\rm H)$ solar value, and the $[\rm O/\rm Fe]$ or alternatively $[\alpha/\rm Fe]$ values as a function of galaxy mass and redshift. These quantities are degenerate in producing the final $[\rm Fe/\rm H]$. 

The solar oxygen abundancy is determined to be $12+\rm log(\rm O/\rm H)=8.66$ (Erb et al. 2006), but other works put it at $12+\rm log(\rm O/\rm H)=8.9$ (as discussed for instance in Liu et al. 2008);
unfortunately, the spread in the adopted value of the solar oxygen abundancy significantly increases the uncertainty in the calculation of the galaxy $\rm M_{star} - [\rm Fe/\rm H]$ relation. 

In lack of direct spectroscopic observations, the determination of $[\rm O/\rm Fe]$ as a function of $[\rm O/\rm H]$, of galaxy mass and of redshift depends on models of both stellar and galaxy evolution, and there is currently no consensus on the conversion $[\rm O/\rm H]$ into $[\rm Fe/\rm H]$ (A. Pipino et al. in preparation, and private communication).
A determination of $[\alpha/\rm Fe]$ as a function of galaxy mass at $\rm z \sim 0$ is provided by Thomas et al. (2005), for a sample of 124 early-type galaxies. The scatter is substantial, and the mass range does not include galaxies below $10^{10} \rm M_{\odot}$. The relation is parameterised as follows: 
\begin{equation}
  [\alpha/\rm Fe]=-0.459+0.062 \ \rm log \rm M_{star} 
\label{thomas1}
\end{equation}

If we use this prescription to convert the $\rm z \sim 0$ Maiolino and Tremonti relations, as $[\rm Fe/\rm H] = [\rm O/\rm H]-[\alpha/\rm Fe]$, we obtain $\rm M_{star} - [\rm Fe/\rm H]$ relations that are consistent (inside the scatter) with the one provided by Thomas et al. (2005) in the range $\rm M_{star} = [10^{10}-10^{12}] \rm M_{\odot}$, provided that $[\rm Fe/\rm H]$ and $[\rm O/\rm Fe]$ are reasonable proxies for $[\rm Z/\rm H]$ and $[\alpha/\rm Fe]$ respectively. The comparison yields values $[\alpha/\rm Fe] \sim 0.1$ for $\rm M_{star} \sim 10^{10} \rm M_{\odot}$ and  $[\alpha/\rm Fe] \sim 0.18$ for $\rm M_{star} \sim 10^{11} \rm M_{\odot}$. 

We use the Maiolino et al. (2008) relations to obtain the $[\rm Fe/\rm H]-\rm M_{star}$ relations at higher redshifts, but we need to make an assumption about the redshift dependence of $[\alpha/\rm Fe]$. Such dependence is very uncertain and not all factors responsible for the variation of $[\alpha/\rm Fe]$ are currently understood; for instance, a progressively top-heavy IMF at higher redshift would cause an excess of oxygen that would speed up the $[\alpha/\rm Fe]$ evolution. For this reason, we choose to calibrate our $[\alpha/\rm Fe]$ $vs$ redshift relation empirically.
Shapiro et al. (2010) use an estimated $[\alpha/\rm Fe] \sim 0.3$ to obtain a relation at redshift $\rm z \sim 2$ from data of $12+\rm log(\rm O/\rm H)$ from Erb et al. (2006). A comparison with the  $\rm z \sim 2$ relation we obtain from Maiolino et al. (2008) via the Thomas et al. (2005) prescription, shows us that we need to assume that $[\alpha/\rm Fe]$ evolves by 0.2 dex in order for the two relations to match. We then extrapolate this evolution linearly with redshift, and obtain values $[\alpha/\rm Fe] \sim (0.1, 0.2, 0.3, 0.5)$ for redshifts $\rm z \sim (0, 1, 2, 3.5)$.

We provide a rough estimate of the error in the $[\rm Fe/\rm H]-\rm M_{star}$  relations from the scatter in the $\rm M_{star}  \ - \ 12+\rm log(\rm O/\rm H)$ relation ($\sim 0.2$ dex, Tremonti et al. 2004), the scatter in $[\alpha/\rm Fe]$ at redshift 0 ($\sim 0.1$ dex, Thomas et al. 2005) and the uncertainty in the solar value of $12+\rm log(\rm O/\rm H)$ ($\sim 0.2$ dex); we obtain an uncertainty $\sigma \sim 0.3$ dex on $[\rm Fe/\rm H]$ for any given stellar mass. Note that this estimate does not take into account the errors in the galaxy mass estimates, nor the error increase in the metallicity measurements at higher redshifts, and nonetheless $\sigma \sim 0.3$ is of the same order of the systematic errors induced by our choice of the $[\alpha/\rm Fe]$ evolution. 
We will explore the consequences of varying these relations in the next Section.  

In Fig.~(\ref{mets}) we plot our fiducial galaxy $[\rm Fe/\rm H]-\rm M_{star}$ relations up to redshift $\rm z \sim3.5$. On the same Figure, we plot the observed relations between the galaxy stellar mass and the GC metallicity $[\rm Fe/\rm H]_{GC}-\rm M_{star}$ obtained by Peng et al. (2006, their Fig. 14). These are shown as the \textit{straigh lines}: \textit{black} for the average GC metallicity in each galaxy, \textit{red} for the metal-rich GCs, and \textit{blue} for the metal-poor GCs.   

Once we have the galaxy $[\rm Fe/\rm H]-\rm M_{star}$ relation in place as a function of redshift, we use it to assign a metallicity to all the globular clusters in the merger tree: the GCs formed in a galaxy of mass $\rm M_{star}$ at a redshift $\rm z$ have a mean metallicity corresponding to the galaxy $[\rm Fe/\rm H]$ at that redshift, according to the derived relations. In addition, in each galaxy the GC metallicity is assumed to have a gaussian distribution around the mean value, with $\sigma \sim 0.3$ dex.

\section{Results}

\begin{figure*} 
\includegraphics[scale=0.8]{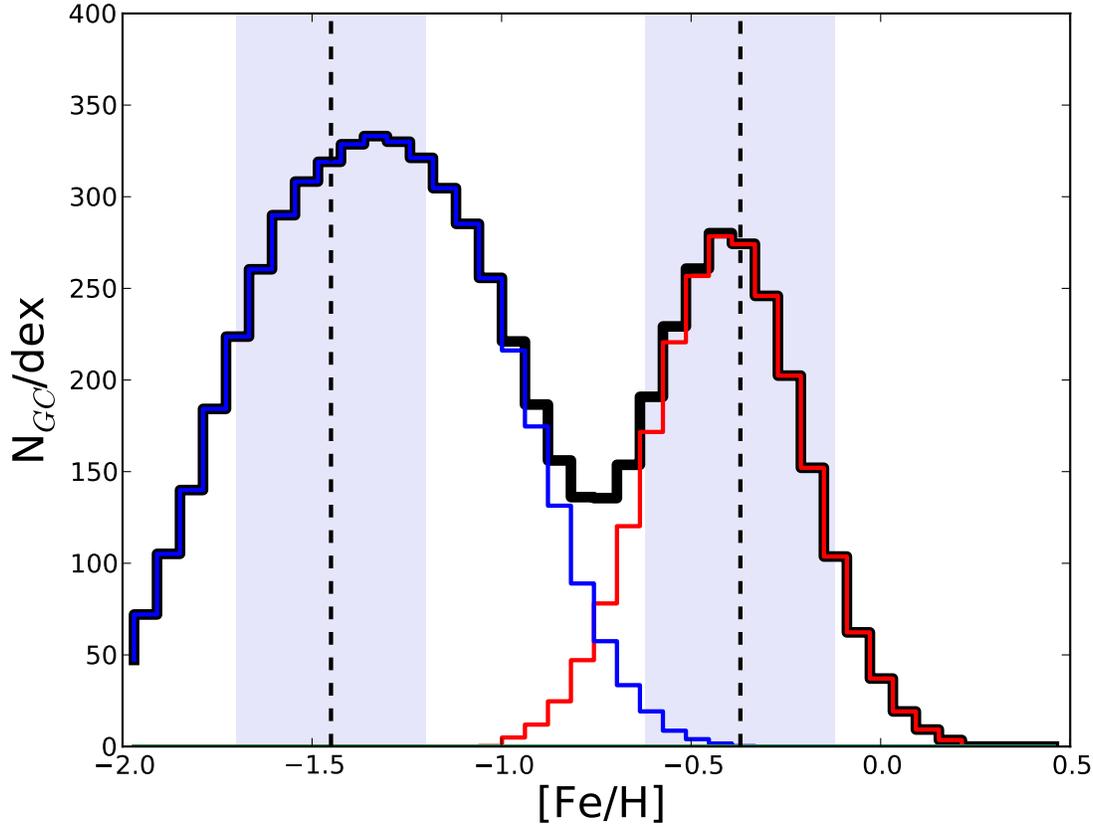}  
\caption{The GC metallicity distribution of a galaxy of stellar mass $\rm M_0 = 10^{11} \ \rm M_{\odot}$. The history of this galaxy is characterised by 
the parameters $\rm M_1/\rm M_0 = 0.3$, $\rm M_{SF}/\rm M_0 = 0$. The local GC metallicity is sampled from a gaussian distribution centered around the galaxy $[\rm Fe/\rm H] - \rm M_{star}$ relation at redshift $\rm z \sim 2$, while the metallicity of GCs accreted from satellites is centered around the galaxy $[\rm Fe/\rm H] - \rm M_{star}$ relation at redshift $\rm z \sim 3.5$.
 \textit{Black line:} total GC metallicity distribution; \textit{red line:} metallicity of local GCs; \textit{blue line:} metallicity of GCs accreted from satellites. \textit{Dotted lines and shaded areas:} values of [Fe/H] for the metal-rich and metal-poor GC populations of a galaxy of $\rm M_{star}=\rm M_0$ from the relations of Peng et al. (2006) and corresponding scatter.}
\label{bimodality}
\end{figure*}

Fig.~(\ref{bimodality}) shows the globular cluster metallicity distribution, in a galaxy of mass $\rm M_0=10^{11} \ \rm M_{\odot}$ at redshift $\rm z=0$, with parameters $\rm M_1/\rm M_0 = 0.3$ and $\rm M_{SF}/\rm M_0 = 0$ (i.e. $70 \%$ of the final stellar mass come from accreted satellites, and there is no additional star formation in the main progenitor after the GC formation), averaged over $\rm N=1000$ Monte Carlo realisations of the galaxy formation history. 
The \textit{thick black line} shows the total distribution, while the \textit{red line} shows the distribution for the clusters that were formed locally in the main progenitor at $\rm z=\rm z_{in} \sim 2$, and the \textit{blue line} shows the distrubution for the clusters formed in satellites at an epoch $\rm z \sim 3-4$, and that merged with the main progenitor. 
The \textit{dashed lines $\rm +$ shaded areas} show the values of the metal-rich and metal-poor GC metallicity $[\rm Fe/\rm H]$ and their $1\sigma$ uncertainties for a galaxy of mass $\rm M_0$ from the Peng et al. (2006) relations. 

The model galaxy shows a sharp bimodality in the globular cluster metallicity distribution. 
The \textbf{metal-rich peak} of the metallicity distibution is entirely dominated by local GCs, formed in the main progenitor at $\rm z \sim 2$. The \textbf{metal-poor peak} is entirely dominated by satellite GCs, accreted via the hierarchical assembly. 
The positions of both peaks are consistent with the observed \textit{galaxy stellar mass - GC metallicity} relations by Peng et al. (2006) for metal-rich and metal-poor globular clusters. 

The number of GCs in various realisations of this galaxy scatters around the value $\rm T_N \sim 6$ interpolated from Peng et al. (2008) for a galaxy of mass $\rm M_{star} \sim 10^{11} \rm M_{\odot}$, staying in the (rather large, $\pm 5$) observed scatter limits. For a given set of history parameters 
$\rm M_1/\rm M_0 $ and $\rm M_{SF}/\rm M_0$, the final value of $\rm T_N$ depends mainly on the assumption about the minimum stellar mass of a galaxy that can form globular clusters (here $\rm M_{limit} = 10^9 \rm M_{\odot}$). Note that a variation of a factor 10 in the mass limit, such that $\rm M_{limit} = 10^8 \rm M_{\odot}$, yields a value $\rm T_N \sim 50$, one order of magnitude off the Peng et al. (2008) relation. 

\begin{figure} 
\includegraphics[scale=0.45]{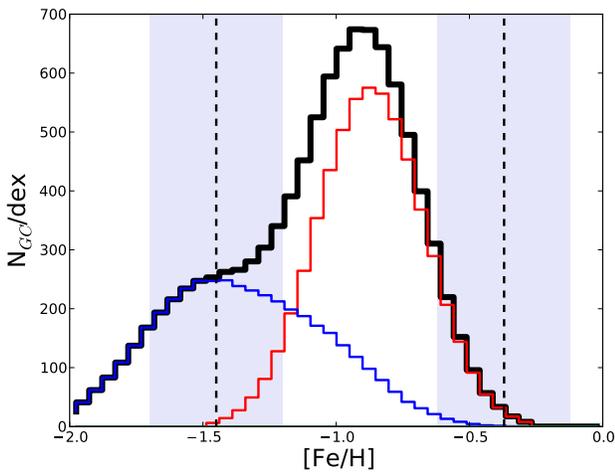}  
\caption{The GC metallicity distribution of a galaxy characterised by ($\rm M_0=1e11 \ \rm M_{\odot}$, $\rm M_1/\rm M_0=0.5$, $\rm M_{SF}=0$), where the GC metallicity of both the local and the accreted GC populations are sampled from a gaussian distribution centered around the average globular cluster $[\rm Fe/\rm H]_{GC} - \rm M_{star}$ relation from Peng et al. (2006: \textit{black line of Fig.~(\ref{mets})}. The \textit{dotted lines and shaded areas}  represent the values of [Fe/H] for the metal-rich and metal-poor GC populations of a galaxy of $\rm M_{star}=\rm M_0$ from the relations of Peng et al. (2006) and corresponding scatter.}
\label{blackline}
\end{figure}

The positions of the peaks in the GC metallicity distribution are determined by the galaxy $[\rm Fe/\rm H]- \rm M_{star}$ relation, given that we can constrain the ages of the GCs from observations, and under the assumption that the GC metallicity is connected to the instantaneous metallicity of the galaxy where they were formed. But is the bimodality just a result of these choices, or is it an intrinsic feature of our mass assembly scenario?
Fig.~(\ref{blackline}) shows the GC metallicity distribution of a galaxy characterised by ($\rm M_0=10^{11} \ \rm M_{\odot}$, $\rm M_1/\rm M_0=0.5$, $\rm M_{SF}=0$) , this time under a very conservative assumption: the GC metallicity in all the objects in the merger tree (main progenitor and satellites) is just taken from the average $[\rm Fe/\rm H]_{GC}-\rm M_{star}$ relation of Peng et al. (2006) (\textit{thick black line} in Fig.\ref{mets}). 

Although both the local and the accreted GCs obey the same average relation, they are still separated in metallicity, the distribution of which shows two distinct peaks, albeit at the wrong values. The reason why the metallicity bimodality is still present is that the hierarchical mass assembly is governed by the halo merger rate, which greatly suppresses merger events of high mass ratios (larger than $1:10$) (as evident in Fakhouri et al. 2010), so that it is highly improbable that a galaxy merges with objects of similar mass, and therefore similar metallicity. This feature alone is what drives the bimodality in the GC metallicity distribution. Therefore, \textbf{a metallicity bimodality in the GC population is a direct prediction of the hierarchical clustering scenario}. 

Notice also that the metal-poor peak in Fig.~(\ref{blackline}) is almost at the right value of $[\rm Fe/\rm H]$, while the metal-rich peak is off by $\sim 0.5$ dex towards the metal-poor side. The slope of the average $[\rm Fe/\rm H]_{GC}-\rm M_{star}$ relation in Fig.\ref{mets} suggests that the number of metal-poor GCs is highly dominant in low-mass galaxies. This happens because these are intrinsically metal-poor galaxies; in addition note that, as their stellar mass is small, in their assembly history they are only able to accrete smaller objects that are devoid of globular clusters (given the existence of $\rm M_{limit}$), therefore their GC population is not bimodal, and their average metallicity peaks exactly where the metal-poor peak is located. On the other hand, the more massive a galaxy is, the richest its assembly history is, with a merger tree with enough mass range to sustain a varied secondary GC population, so its GC population is more likely to be bimodal.
Therefore, a massive galaxy always has a secondary, metal-poor GC population, and the \textit{average} GC metallicity deviates from both peaks. This point is addressed in the next Figure. 

\begin{figure} 
\includegraphics[scale=0.45]{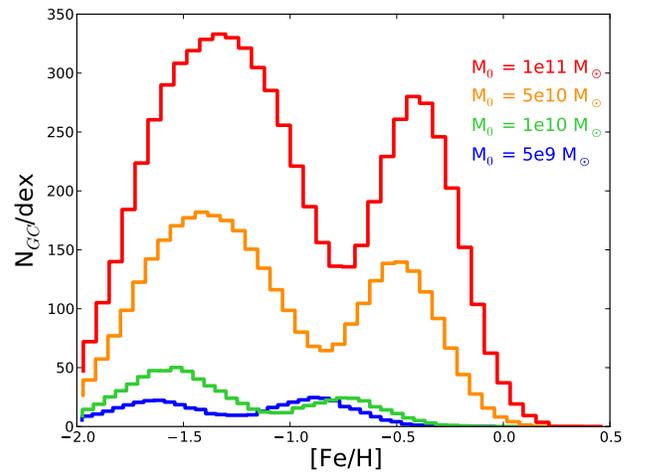}    
\caption{The total GC metallicity distribution of galaxies of varying final stellar mass, all characterised by the parameters $\rm M_1/\rm M_0 = 0.3$, $\rm M_{SF}/\rm M_0 = 0$.}
\label{masses}
\end{figure}

Fig.~(\ref{masses}) illustrates the difference in the GC total metallicity distribution of galaxies with final stellar masses $\rm M_0=10^{11}, 5 \times 10^{10}, 10^{10}, 5 \times 10^9 \ \rm M_{\odot}$, all characterised by the history parameters $\rm M_1/\rm M_0 = 0.3$, $\rm M_{SF}/\rm M_0 = 0$. 
The bimodality in the GC metallicity distribution is evident at all masses in this mass range. 
As expected, the more massive a galaxy is, the richest is its GC population, in both the metal-rich and the metal-poor component.
However, notice that, although the history parameters are the same in all cases, the relative contribution of the two peaks varies, with the metal-poor peak becoming less and less significant relative to the metal-rich peak for lower-mass galaxies, in accord with Peng et al. (2008) and Shapiro et al. (2010).
In the lowest mass bin, the relative height of the metal-rich and metal-poor peaks is reversed; given that the mass limit for GC formation is $10^9 \ \rm M_{\odot}$, this galaxy is for the most part accreting satellites that don't contribute to the GC population, with the rare exception of major mergers (in this case, $\rm M_{sat} > 10^9 \ \rm M_{\odot}$). If we assume that the globular cluster formation is hampered in low-mass galaxies, i.e. that galaxies below a mass threshold cannot produce globular clusters, then the model predicts that the GC metallicity bimodality ceases to exist slightly above that mass threshold. In such galaxies, the GC population is unimodal and entirely composed of locally-formed GCs. On the other hand, the overall GC metallicity decreases following the galaxy mass, and as a result, the GC population in low mass galaxies is metal-poor, again in accord with Peng et al. (2008). 

\begin{figure*} 
\includegraphics[scale=0.37]{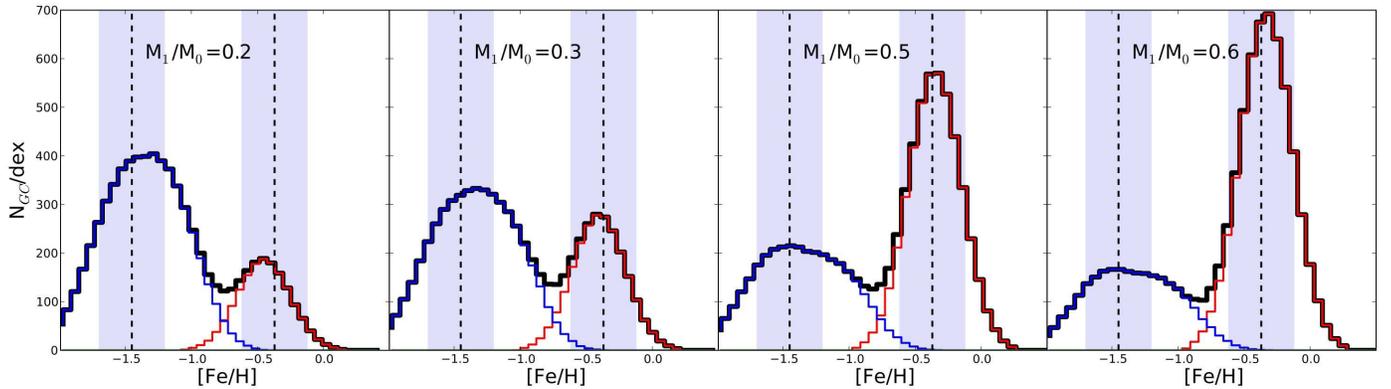}  
\caption{The variation of the GC metallicity distribution for different assembly histories, parameterised by $\rm M_1/\rm M_0$. \textit{From left to right:} $\rm M_1/\rm M_0 = (0.2, 0.3, 0.5, 0.6)$. In all cases, the final galaxy stellar mass is $\rm M_0=10^{11} \ \rm M_{\odot}$, and $\rm M_{SF}=0$. The \textit{dotted lines and shaded areas}  represent the values of [Fe/H] for the metal-rich and metal-poor GC populations of a galaxy of $\rm M_{star}=\rm M_0$ from the relations of Peng et al. (2006) and corresponding scatter.}
\label{m1m0}
\end{figure*}

The galaxy assembly history determines the fraction of the final mass that is accreted from the merger tree, and therefore the fraction of globular clusters that are formed outside the main progenitor and which we have shown to compose the metal-poor peak.
Fig.~(\ref{m1m0}) shows the relative height of the metal-rich and metal-poor peaks generated in different assembly histories, parameterised as $\rm M_1/\rm M_0$. For a galaxy of final stellar mass $\rm M_0=10^{11} \ \rm M_{\odot}$, the \textit{panels from left to right } show the GC metallicity distribution for $\rm M_1/\rm M_0 = (0.2, 0.3, 0.5, 0.6)$ respectively. In all cases, $\rm M_{SF}=0$. As expected, a galaxy with a poor merger history (such as the case $\rm M_1/\rm M_0=0.6$ for instance) shows a GC metallicity distribution dominated by the local metal-rich population.
The model therefore predicts that the presence of a very strong metal-rich GC component is a sign of a sparse merger history. For a given galaxy mass, the richness of the merger tree depends on enviromnent; hence the model predicts that galaxies in low-density environments have, for a given mass, a GC population that is more metal-rich dominated than galaxies of the same mass living in the centre of clusters. 

\begin{figure*} 
\includegraphics[scale=0.37]{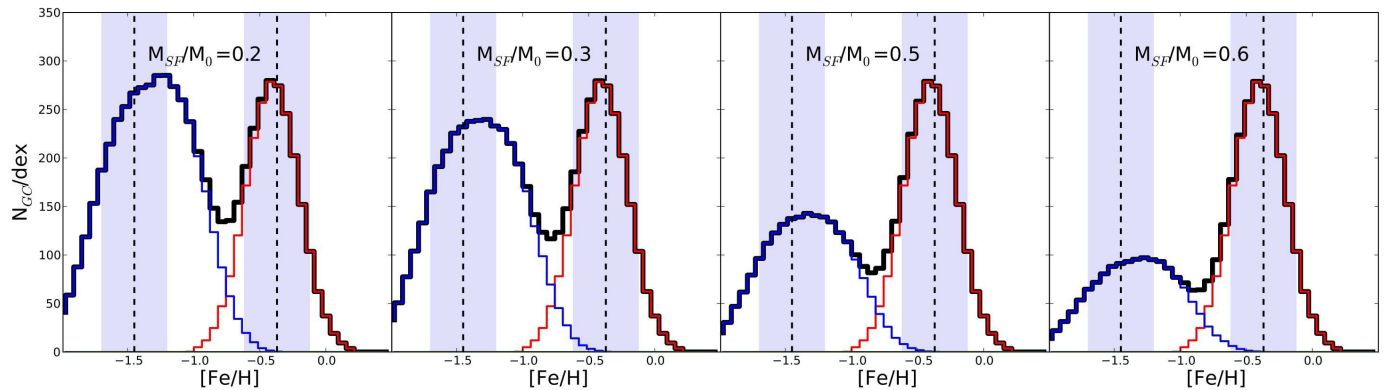}  
\caption{The variation of the GC metallicity distribution for different star formation histories, parameterised by $\rm M_{SF}/\rm M_0$. \textit{From left to right:} $\rm M_{SF}/\rm M_0 = (0.2, 0.3, 0.5, 0.6)$. In all cases, the final galaxy stellar mass is $\rm M_0=10^{11} \ \rm M_{\odot}$, and $\rm M_1/\rm M_0=0.3$. The \textit{dotted lines and shaded areas}  represent the values of $[\rm Fe/\rm H]$ for the metal-rich and metal-poor GC populations of a galaxy of $\rm M_{star}=\rm M_0$ from the relations of Peng et al. (2006) and corresponding scatter. }
\label{SF}
\end{figure*}

So far we have analysed the simplified case of galaxies with $\rm M_{SF}=0$. However,
for the majority of galaxies the star formation does not stop at $\rm z \sim 2$, and a significant part of the final stellar mass is formed at later times. In this case, a significant fraction of the galaxy stellar mass is not associated with formation or accretion of globular clusters.
To account for this stellar component, we vary the value of the star formation history parameter $\rm M_{SF}/\rm M_0$. 
Fig.~(\ref{SF}) shows the GC metallicity distribution for a galaxy of final mass $\rm M_0=10^{11} \ \rm M_{\odot}$, where the stellar mass is contributed by $1)$ the main progenitor at the epoch of GC formation in proportion of $\rm M_1/\rm M_0 = 0.3$ (the local GC population is associated with this component), $2)$ stars formed locally in the galaxy \textit{after} the epoch of GC formation, in quantity $\rm M_{SF}/\rm M_0 = (0.2, 0.3, 0.5, 0.6)$ (\textit{panels from left to right}), and $3)$ stars accreted from satellites, in quantity $\rm M_0-\rm M_1-\rm M_{SF}=\rm M_{sat}$. Note that a higher value of $\rm M_{SF}/\rm M_0$ implies a smaller value of $\rm M_{sat}$, i.e. a poorer merger history. 

Fig.~(\ref{SF}) shows that, as the galaxy growth becomes more dominated by local star formation and the contribution of the mass accreted by satellites is smaller, 
the GC population becomes more and more dominated by local globular clusters, even if most of the stellar component is not directly associated with the globular clusters themselves. The model predicts that, in galaxies with an active star formation history \textit{after} the GC formation (i.e. at $\rm z<2$), the relative strenght of the metal-rich and metal-poor peaks of the GC metallicity distribution is biased towards the metal-rich GCs, for a given galaxy mass.
If we consider the star formation history as associated with morphology, then the model predicts that, for a given galaxy mass $\rm M_0$ and total number of GCs, late-type galaxies have a GC metallicity distribution with a stronger metal-rich peak than early-type galaxies. 
 
Note that the results in Figs.~(\ref{m1m0}, \ref{SF}) show that, given the final mass of the galaxy $\rm M_0$, the final number of globular clusters in the galaxy depends on the value of the assembly history parameter $\rm M_1/\rm M_0$ and the star formation history parameter $\rm M_{SF}/\rm M_0$. The large scatter in the value of the GC frequency per unit mass $\rm T_N$ for a given galaxy mass seen in Peng et al. (2008) is likely to be due to the variety of histories for galaxies in each mass bin. The final value of $\rm T_N$ decreases for a decreasing value of $\rm M_{sat}$.
 Note that the assembly and star formation history parameters have instead no effect on the position of the peaks, which are entirely determined by the evolution of the galaxy $[\rm Fe/\rm H]-\rm M_{star}$ relation.
The number of GCs, together with the relative abundance of the metal-rich and metal-poor components, can therefore be used to constrain the assembly and star formation history of the galaxy.

\begin{figure} 
\includegraphics[scale=0.45]{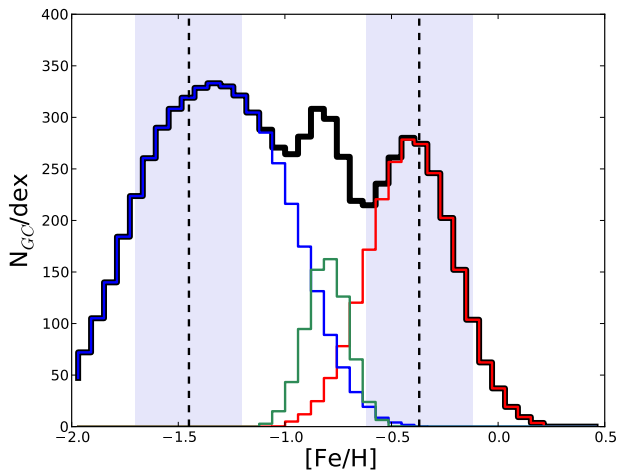}    
\caption{The GC metallicity distribution of a galaxy of mass $\rm M_0=10^{11} \ \rm M_{\odot}$ with $\rm M_1/\rm M_0=0.3$ (see Fig.~(\ref{bimodality})), with the inclusion of a  major gas-rich merger event in the assembly history. The merger triggers the formation of a tertiary population of GCs, in number $\rm N=0.3 \times \rm N(\rm M_1)$ (i.e. $30 \%$ of the main progenitor's local GCs) and intermediate metallicities peaked around $[\rm Fe/\rm H] \sim -0.8$ (\textit{green line}).}
\label{merger}
\end{figure}

As discussed in the Introduction, the scenario in which globular cluster are generally formed in gas-rich mergers cannot reproduce the properties and the scaling relations of the GC population. However, if globular clusters indeed form during violent bursts of star formation, it is physically possible that \textit{some} of them indeed are formed in mergers at all redshifts, a fact that would explain the presence of intermediate-age or young GCs in some galaxies (Kissler-Patig et al. 1998, Puzia et al. 2005, Strader et al. 2003, 2004b, Woodley et al. 2010, Brodie $\&$ Strader 2006 and references therein).
Fig.~(\ref{merger}) shows the effect of a gas-rich merger event, where new GCs are formed, on the GC metallicity distribution. In this example, in the same galaxy portrayed in Fig.~(\ref{bimodality}), we introduce a gas-rich merger event that triggers the formation new GCs, in quantity $\sim 30 \%$ of the local GC population of the main progenitor,
with intermediate metallicities peaked around $[\rm Fe/\rm H] \sim -0.8$ with a gaussian of width $\sigma = 0.2 \ \rm dex$ (\textit{green line}). This plot shows that the creation of new GCs in merger events introduces a stochastic variation of the GC metallicity distribution, that leads to the formation of tertiary peaks, in positions determined by the metallicity of the gas perturbed/carried by the merger. The number of newly-formed GCs depends on the available gas mass and the star formation rate in the merger-triggered bursts, as well as the efficiency of GC formation $vs$ star formation. 

It is clear from this plot that, if we consider the formation of GC in gas-rich merger events, the GC metallicity distribution becomes more complex. The stochasticity of such events allows for any shape of the final metallicity distribution: a prolongued history of gas-rich mergers contributes to the dilution of the bimodality. Such mechanism can explain the number of 'exotic' GC metallicities distributions found by a number of authors, including Blom et al. (2012) and Peng et al. (2006), 
with a number of galaxies that either show one or more tertiary peaks, or a non-bimodal GC metallicity distribution 
(it should be noted that such a scenario needs to be confirmed with dynamical analysis; for instance, Blom et al. 2012 show data of a galaxy with an intermediate-metallicity GC subpopulation that rotates with the main body of the galaxy). 
Note that major gas-rich mergers are good candidates to provide very intense bursts of star formation, during which new globular clusters can be formed. If a galaxy undergoes an assembly history devoid of any gas-rich mergers, it is hard to envisage another mechanism that is able to provide a strong enough perturbation of the gas in the galaxy to trigger very intense bursts of star formation (as per Shapiro et al. 2010), which can create a tertiary GC population. Therefore, we can consider tertiary peaks in the GC metallicity distribution as clear signatures of major gas-rich merger events in the past history of the galaxy. 

\begin{figure*} 
\includegraphics[scale=0.8]{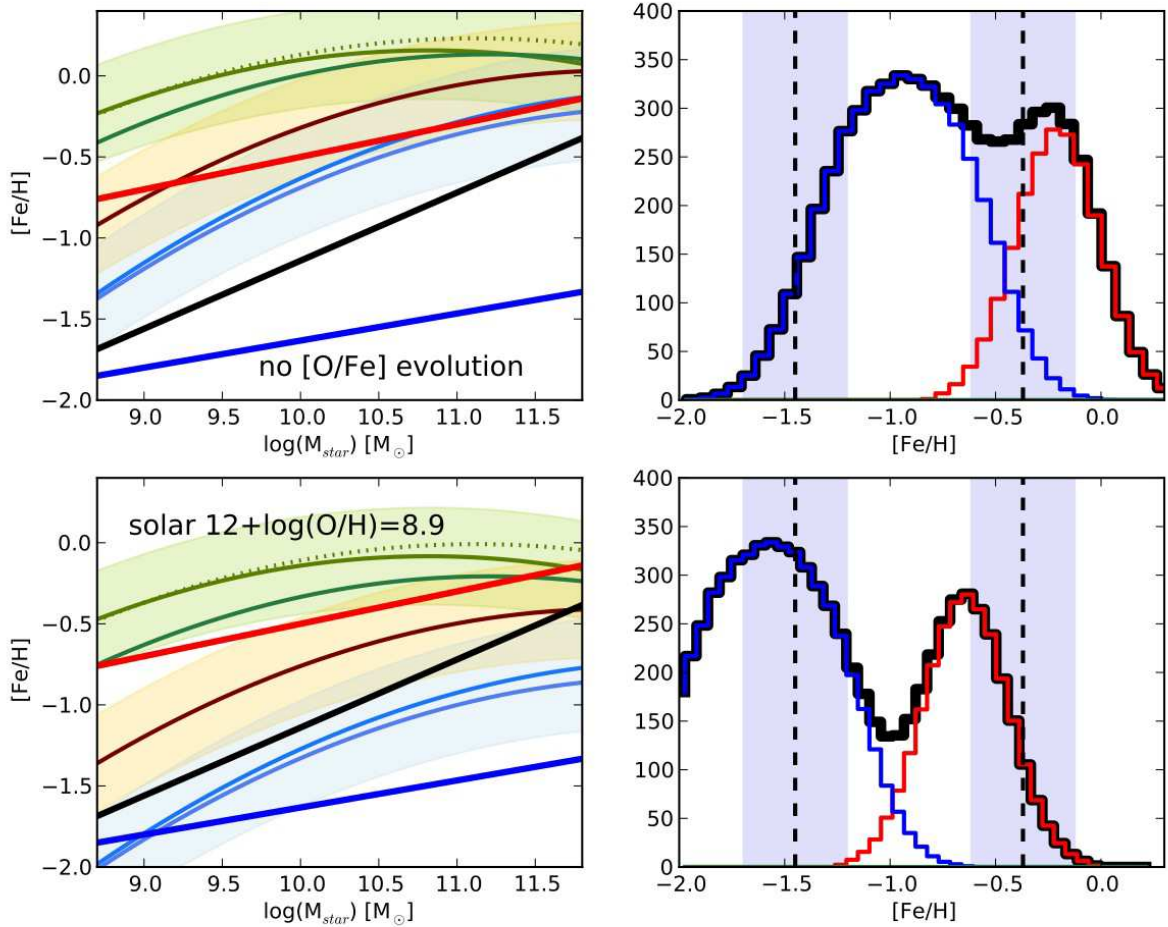}    
\caption{The dependence of the GC metallicity distribution on the evolution of the galaxy mass-metallicity relation. \textit{Upper left panel:} the galaxy $[\rm Fe/\rm H] - \rm M_{star}$ relation as a function of redshift (as per Section 3, compare with Fig.~(\ref{mets})), in the case of no $[\rm O/\rm Fe]$ evolution. \textit{Upper right panel:} the corresponding GC metallicity distribution. \textit{Lower left panel:} the galaxy $[\rm Fe/\rm H] - \rm M_{star}$ relation with the $[\rm O/\rm Fe]$ evolution described in Section 3, but with the oxygen solar value set as $12+\rm log(\rm O/\rm H)=8.9$. \textit{Lower left panel:} the corresponding GC metallicity distribution. 
In the \textit{right-hand side panels,} the portrayed galaxy is characterised by ($\rm M_0=10^{11} \ \rm M_{\odot}$, $\rm M_1/\rm M_0=0.3$, $\rm M_{SF}=0$); the \textit{dotted lines and shaded areas}  represent the values of [Fe/H] for the metal-rich and metal-poor GC populations of a galaxy of $\rm M_{star}=\rm M_0$ from the relations of Peng et al. (2006) and corresponding scatter.}
\label{gcmets}
\end{figure*}

The total number of GCs and the relative height of the metal-rich and metal-poor peaks depend on the galaxy mass and the galaxy assembly and star formation history, while the positions of the peaks depend on the determination of the galaxy mass-metallicity relation as a function of redshift. To explore this point further, we study the GC metallicity distribution resulting from the $[\rm Fe/\rm H] - \rm M_{star}$ relation obtained in Section 3, under different prescriptions. In particular, we describe two examples of variations of the recipes described in Section 3, that affect 1) the \textit{normalisation} and 2) the \textit{evolution} of the galaxy $[\rm Fe/\rm H] - \rm M_{star}$ relation: 1) is a variation of the 
value for the solar $12+\rm log(\rm O/\rm H)$, and 2) is a variation of the prescription for the evolution of the galaxy $[\alpha/\rm Fe]$ value.

Fig.~(\ref{gcmets}) shows the GC metallicity distribution of a galaxy characterised by ($\rm M_0=10^{11} \ \rm M_{\odot}$, $\rm M_1/\rm M_0=0.3$, $\rm M_{SF}=0$) (\textit{right panels}) resulting from two different sets of galaxy $[\rm Fe/\rm H] - \rm M_{star}$ relations as a function of redshift (\textit{left panels}), which are derived following the model of Section 3. In the \textit{upper left panel}, the galaxy $[\rm Fe/\rm H] - \rm M_{star}$ relation is obtained from the evolution of $12+log(\rm O/\rm H)$, but assuming that the galaxy $[\rm O/\rm Fe]$ does not evolve with redshift, but has a constant value of $\sim 0.1$ like in the local universe (with a solar oxygen value $12+\rm log(\rm O/\rm H)=8.66$ as in our fiducial relation).
In the \textit{upper right panel}, the resulting galaxy GC metallicity distribution shows a somewhat diminished bimodality, and the positions of the peaks are definitely off the observed values obtained by Peng et al. (2006; \textit{dotted lines and shaded areas}). Both peaks are centered around too high metallicities, and the problem is worse for the metal-poor peak; if we assume that the galaxy $[\rm O/\rm Fe]$ does not evolve with redshift but mantains the local value, then we are overestimating the galaxy metal content by a factor that is proportional with redshift, and the globular clusters formed in small galaxies at high redshifts are the ones that are affected the most. 

The \textit{lower left panel} of Fig.~(\ref{gcmets}) portrays the galaxy $[\rm Fe/\rm H] - \rm M_{star}$ relations as obtained in Section 3, with the same evolution of $[\rm O/\rm Fe]$ used so far in this work, but with a different solar oxygen value, $12+\rm log(\rm O/\rm H)=8.9$ (Liu et al. 2008). In this case, the evolution of the $[\rm Fe/\rm H] - \rm M_{star}$ relation is not affected, but its normalisation has changed. The resulting GC metallicity distribution does not change in shape (the relative height and position of the peaks is the same) but both peaks are shifted towards lower $[\rm Fe/\rm H]$ values, because the oxygen content $[\rm O/\rm H]$ yields a lower total $[\rm Fe/\rm H]$ content.

The model predicts that the positions of the metal-rich and metal-poor peaks of the GC metallicity distribution are exclusively dependent on the galaxy mass-metallicity relation as a function of redshift. In the case where globular clusters are indeed fossil records of the metallicity of their parent galaxy at the time when they formed, then through this model the GC metallicity distribution, and in particular the position of the metal-rich and metal-poor peaks, can be used to constraint the evolution of the galaxy $[\rm Fe/\rm H] - \rm M_{star}$ relation and the evolution of the galaxy $[\rm O/\rm Fe]$.

\section{Discussion}   

The results described in the previous Section show that the hierarchical galaxy assembly directly predicts a metallicity bimodality of the globular cluster populations in galaxies. Previous studies like Shapiro et al. (2010) on the contrary, argued that the hierarchical mass assembly would blur the correlations between galaxies and their GCs.   
The model presented here predicts globular clusters to have a bimodal metallicity distribution, with the metal-rich and the metal-poor peaks following the $[\rm Fe/\rm H]_{GC} - \rm M_{star} $ relations described in Peng et al. (2006), and determined by the evolution of the galaxy mass-metallicity relation. The model also predicts that the relative strenght of the metal-rich and metal-poor peaks depends on the assembly and star formation history of the galaxy. Moreover, the model predicts that the metallicity bimodality disappears for masses below $\rm M_{star} \sim 10^9 \rm M_{\odot}$ and for redshifts $\rm z>2$. 

The main mechanism at work in producing the GC bimodality is the existence of the galaxy mass-metallicity relation, coupled with the natural behaviour of hierarchical mass assembly to strongly favour minor mergers over major mergers. In this scenario, all globular clusters form with the same mechanism and share the metallicity of their parent galaxy; the hierarchical build-up of galaxies then assembles the GC populations so that the GCs of a satellite become part of the accreted, metal-poor GC component of a bigger galaxy of higher metallicity (in a generalisation of the scenario of Shapiro et al. 2010). Note that this mechanism still works in the case of satellites with bimodal GC distributions: both the metal-rich and the metal-poor subpopulations of a satellite are metal-poor compared to the central galaxy, and will contribute to its final metal-poor GC population.

Under the assumption that GC formation is a rapid process, closely linked to the properties of the galaxy at the time of the event, the
fact that GCs of different metallicities form in different galaxies, at slightly different times, is enough for the hierarchical galaxy formation scenario to naturally produce the correct GC metallicity distribution. 
The observational facts imply this very clearly. For instance, a comparison of the metal-poor globular cluster $[\rm Fe/\rm H]_{GC}-\rm M_{star}$ relation with the galaxy mass-metallicity relation (\textit{blue straight line} in Fig.~(\ref{mets})) shows that, for any given galaxy mass above $\rm M_{star} > 10^9 \ \rm M_{\odot}$, the galaxy is always more metal-rich that its own metal-poor globular clusters (unless they were formed at $\rm z>4$). This implies that these globular clusters must have formed in smaller galaxies and they were then accreted. Note that the alternative scenario of a multi-phase GC formation in each single galaxy (like for instance Beasley et al. 2002), where metal-poor GCs are formed first and metal-rich GCs are formed later in mergers, necessarily implies some form of systematic segregations of metals and an \textit{ad hoc} shut-down of the formation of the metal-poor component, in order to produce both the correct metallicity and the bimodality. In fact, if a galaxy experiences a prolongued phase of local GC formation, the bimodality is destroyed by the galaxy chemical self-enrichment.

The results presented in this work are based on theoretical merger trees extracted from the Millennium simulation. Among the cosmological parameters, the value of $\sigma_8$ can affect these results, in the sense that a lower clustering power would produce a delayed mass accretion and sparser merger histories. While the bimodality would remain unchanged, the height of the metal-poor peak
would be affected. However, the magnitude of the effect would be smaller than the scatter in the observed values of $T_N$ and the scatter between different Monte Carlo runs, and the results presented in this work would remain unaffected. In any case, this would be an interesting avenue of investigation.

The bimodality in the GC metallicity distribution is solid against variations of our initial conditions, such as any assumption about the metallicity we assume for the galaxy or the GC themselves (Figs.~\ref{bimodality}, \ref{blackline}, \ref{gcmets}). Thorugh hierarchical galaxy assembly, it is rather hard to produce a GC metallicity distribution that is not bimodal. 
Nonetheless, the position of the peaks is not recovered correctly if we assume that the local and accreted GC populations form at the same time, given the galaxy mass-metallicity relations described in this work. This is in accord with observations that determine the metal-poor GCs to be $1-2$ Gyr older than the metal-rich GCs. A second-order sophistication of this model would be to introduce an analytic relation between the galaxy mass and the epoch GC formation, but in reality this epoch is likely to vary with environment and the fluctuations of the local star formation rate. A scatter in the epoch of formation would mirror a scatter in the galaxy metallicity, via the evolution of the galaxy mass-metallicity relation; this effect has been mimicked in this work by the introduction of a scatter in the GC metallicity of the satellites. 

In this model, old globular clusters are considered fossil records of the galaxy where they were formed, and their finally metallicity distribution is exclusively a result of the hierarchical galaxy assembly. The factors that affect the final metallicity distributions are 1) the evolution of the galaxy mass-metallicity relation, which \textit{completely} determines the position of the metal-rich and metal-poor peaks, and 2) the merger history (therefore the environmental density) and the star formation history, which \textit{completely} determine the relative strenghts of the metal-poor and metal-rich peaks. These constraints are \textit{independent and orthogonal}, with no degeneracy between them. 

Provided we know the GC ages, we can use this model to test and constrain the evolution of the galaxy $[\rm Fe/\rm H]-\rm M_{star}$ relation and the evolution of $[\alpha/\rm Fe]$, through the positions of the metal-rich and metal-poor peaks. In order to reproduce the Peng et al. (2006) $[\rm Fe/\rm H]_{GC}-\rm M_{star}$ relations, the model favours a value of $[\alpha/\rm Fe]=0.5$ at redshift $\rm z=4$ and a linear evolution down to $[\alpha/\rm Fe]=0.1$ at $\rm z=0$, a solar value $12+\rm log(\rm O/\rm H) \sim 8.66$, and the evolution of the galaxy $[\rm Fe/\rm H]-\rm M_{star}$ relation plotted in Fig.~(\ref{mets}). 
Note however that, even if we can rely on spectroscopy for the determination of the globular cluster metallicity with good precision, there are still significant uncertainties on the GC age determination. In this work we have used the observed estimates for the average GC ages, but if we consider 
their uncertainties, combined with the current size of the uncertaintiy on the galaxy mass-metallicity relation, then 
there is a substantial degeneracy between redshift and galaxy metallicity in the determination of the positions of the metal-rich and metal-poor peaks, an issue that will be solved with higher precision observations from the next-generation instruments and surveys. 

Unrelated to the particular values of the metal-rich and metal-poor peaks, the model predicts the relative height of the peaks to give an insight into the assembly and star formation history of the galaxy: for a given galaxy mass, a dominant metal-rich peak indicates a quiet merger history and/or a significant growth of the galaxy through local star formation, as opposed to a mass growth driven by accretion of satellites. Therefore, the relative abundance of metal-poor and metal-rich GCs is correlated with environmental density and morphology, with isolated late-type galaxies of given mass $\rm M_0$ being more metal-rich GC dominated than early types of the same mass living in dense environments. The current state of observations of GC populations in late-type galaxies does not yet provide definitive constraints, but this is certainly an area where more numerous and improved observations are called for.
This is of particular interest when attempting to constrain the galaxy star formation history, in that this method is complementary to the SED-fitting technique, which is affected by a number of systematic uncertainties (Tonini et al. 2009, 2010, 2011, 2012, Henriques et al. 2011, Maraston et al. 2010, Pforr et al. 2012).

The hierarchical galaxy mass assembly naturally leads to bimodality in the globular cluster metallicity distribution. A bimodal distribution is sign of a two-phase galaxy formation, with an intense dissipative phase that leads to the formation of the core of the galaxy and the local GCs, and a second phase of accretion of satellites and the acquisition of a secondary GC population. This regularity breaks down at very small masses, when galaxies are not massive enough to accrete GCs from their merger tree: their satellites are in fact so small, that they cannot form their own GCs. In this case the GC distribution is unimodal. On the other hand, the smaller a galaxy is, the least probable it is that it actually has an extended merger tree, so even when the few satellites contribute with globular clusters, the metal-poor peak is subdominant or negligible.  
Note that this behaviour is not in disagreement with the $\rm T_N-\rm M_{star}$ relation presented in Peng et al. (2008). For masses below $10^9 \ \rm M_{\odot}$, even 1 globular cluster will yield a value $\rm T_N \sim 10$. This shows that our limit mass for the production of globular clusters, $\rm M_{limit} \sim 10^9$, is a realistic prediction. 

In relation to the number of globular clusters per galaxy mass $\rm T_N$ predicted by the model, the values we obtain are in the range of the scatter observed by Peng et al. (2008).
We argue that in observations such scatter arises from the variety of assembly and star formation histories that generated the galaxies in the sample. In the model, the scatter between different Monte Carlo runs is of the same order of magnitude, and we argue that such a scatter generates from the scatter in the mass function of the merger tree, incorporating all the possible assembly paths to build up the final galaxy mass. 
The final value of $T_N$ depends on the assembly and history parameters $\rm M_1/\rm M_0$ and $\rm M_{SF}/\rm M_0$, where richer merger histories strenghten the metal-poor peak and raise the total number of globular clusters. 

The model also predicts that the GC bimodality is a function of redshift. The more time a galaxy has to accrete satellites after the GC formation, the more rich its secondary population will be. Therefore, we expect the GC metallicity bimodality to disappear by redshift $\rm z \sim 2$ and above.

\section{Summary and conclusions}

We have performed a series of Monte-Carlo simulations of the assembly history of galaxies and studied the metallicity distribution of their globular cluster systems. To do so, we built a model for the assembly of the globular cluster population following the hierarchical galaxy assembly, based on empirical scaling relations such as the galaxy mass-metallicity relation $[\rm O/\rm H]-\rm M_{star}$, and on the observed galaxy stellar mass function up to redshift $\rm z \sim 4$. We also made use of the theoretical merger rates as a function of mass and redshift from the Millennium simulation, to build  merger trees for a set of final galaxies. 

By determining the galaxy $[\rm Fe/\rm H]-\rm M_{star}$ relation for all galaxies in each merger tree, and by assuming that globular clusters share the metallicity of their original parent galaxy at the time of their formation, we populated the merger tree with globular clusters. The hierarchical assembly of the final galaxy creates a globular cluster population composed by the local GCs formed in the main progenitor and those accreted from the merger tree. We conclude that: 

$\bullet$ the final globular cluster metallicity distribution is in general bimodal; the GC metallicity bimodality is a direct prediction of the hierarchical clustering scenario; 

$\bullet$ the metal-rich peak of the GC metallicity distribution is composed of globular clusters locally formed in the main progenitor, while the metal-poor peak is composed of globular clusters accreted from the satellites that compose the merger tree. At all times GCs in satellites are more metal-poor than GCs formed in the main progenitor due to the existence of the galaxy mass-metallicity relation; both the metal-rich and the metal-poor subpopulations of a satellite will contribute to the metal-poor GC population of the main galaxy;

$\bullet$ the positions of the metal-rich and metal-poor peak depend exclusively on the evolution of the galaxy mass-metallicity relation $[\rm Fe/\rm H]-\rm M_{star}$; we are able to constrain such evolution and predict that the galaxy $[\rm O/\rm Fe]$ evolves linearly with redshift from a value of $\sim 0.5$ at redshift $\rm z \sim 4$ to a value of $\sim 0.1$ at $\rm z=0$;

$\bullet$ the relative strenght of the metal-rich and metal-poor peak depends on the assembly and star formation history of the galaxy. The model predicts that, for a given galaxy mass, galaxies with a poor merger history, such as galaxies forming in low density environments, and/or galaxies with a prolongued star formation history (after the epoch of GC formation) that contributes most of the galaxy mass, such as late-type galaxies, will have a globular cluster population dominated by the metal-rich component. On the other hand, galaxies of the same mass but with an intense merger history, such as early-type galaxies and/or galaxies living in dense environments, will have a globular cluster population with a larger metal-poor component; 

$\bullet$ the model predicts that the globular cluster metallicity bimodality disappears at galaxy masses around $10^9 \ \rm M_{\odot}$; moreover, the model predicts that the bimodality is progressively less pronounced at higher redshift, and disappears around redshift $\rm z \sim 2$.

\section*{Acknowledgements} 

CT would like to thank the anonymous Referee for her/his comments and suggestions, which improved the clarity of this work. CT would also like to thank Jeremy Mould, Duncan Forbes, Eric Peng, Lee Spitler, Marie Martig, Vincenzo Pota, Chris Usher and Darren Croton for the interesting discussions and for their useful comments.

\end{document}